# Measuring the frequency of a Sr optical lattice clock using a 120-km coherent optical transfer

F.-L. Hong[1,4,*], M. Musha[2], M. Takamoto[3,4], H. Inaba[1,4], S. Yanagimachi[1], A. Takamizawa[1], K. Watabe[1], T. Ikegami[1], M. Imae[1,4], Y. Fujii[1,4], M. Amemiya[1,4], K. Nakagawa[2], K. Ueda[2], and H. Katori[3,4]

[1]*National Metrology Institute of Japan (NMIJ), National Institute of Advanced Industrial Science and Technologies (AIST), Tsukuba, Ibaraki 305-8563, Japan*

[2]*Institute for Laser Science, University of Electro-communications, 1-5-1 Chofugaoka Chofu-shi, Tokyo 182-8585, Japan*

[3]*Department of Applied Physics, Graduate School of Engineering, The University of Tokyo, Bunkyo-ku, Tokyo 113-8656, Japan*

[4]*CREST, Japan Science and Technology Agency, 4-1-8 Honcho Kawaguchi, Saitama 332-0012, Japan*

[*]*Corresponding author: f.hong@aist.go.jp*

We demonstrate a precision frequency measurement using a phase-stabilized 120-km optical fiber link over a physical distance of 50 km. The transition frequency of the $^{87}$Sr optical lattice clock at the University of Tokyo is measured to be 429228004229874.1(2.4) Hz referenced to international atomic time (TAI). The measured frequency agrees with results obtained in Boulder and Paris at a $6\times10^{-16}$ fractional level, which matches the current best evaluations of Cs primary frequency standards. The results demonstrate the excellent functions of the intercity optical fibre link, and the great potential of optical lattice clocks for use in the redefinition of the second.

The precision measurement of time and frequency is of great interest for a wide range of applications including fundamental science, metrology, broadband communication networks and navigation with global positioning systems (GPS). The development of optical frequency measurement based on femtosecond combs [1, 2] has stimulated the field of frequency metrology, especially research on optical frequency standards. An 'optical lattice clock' was proposed recently [3] in which atoms trapped in an optical lattice serve as quantum references. The light shift induced by the trapping field can be precisely cancelled out by carefully designing the light shift potential, making it possible to achieve a frequency uncertainty below $10^{-17}$ [4]. By late 2006, frequency measurements for Sr lattice clocks had been carried out by groups at Tokyo-NMIJ [4, 5], JILA (Boulder) [6], and SYRTE (Paris) [7], and the agreement between the results obtained by three groups reached the $7.5\times10^{-15}$ fractional level. Based on these results, the Sr lattice clock was recommended by the International Committee for Weights and Measures (CIPM) as one of 'the secondary representations of the second' (candidates for the redefinition of the second).

One major challenge for scientists working on high-precision frequency standards and measurements is to deliver and compare state-of-the-art clocks in different locations. The ubiquitous fiber optic network has the potential to allow us to transfer frequencies with extremely low phase noise [8-14]. Microwave frequencies have been faithfully transmitted over a modulated optical signal with a fractional frequency instability of $8\times10^{-15}$ at 1 s through a 43-km fiber [10]. By using the higher frequencies available with an optical carrier, we can further increase the signal counting rate and achieve an instability that is several orders of magnitude lower [9]. This very stable optical carrier transfer can be used to compare two optical clocks without degrading their stability and accuracy [14]. Results have been reported for tests of coherent optical frequency transfer with residual frequency instabilities of $1\times10^{-17}$ through a 32-km fiber [11] and $2\times10^{-16}$ through 251-km fibers [12] at 1 s. In proof-of-principle studies for distant optical transfer, most of the fibers consisted of a stable exclusive fiber network including spooled fibers in the laboratory, whose loss and intrinsic noise are relatively small compared with actually installed fiber networks. Here, we demonstrate the first precision frequency measurement based on an optical carrier transfer over a 120-km long-haul fiber network composed of installed rural and urban telecom fibers. The transition frequency of the Sr optical lattice clock in Tokyo is determined in as little as 3 hours, with an uncertainty about half that in our previous 9-day measurement using a GPS link [5].

Figure 1 is a diagram of a remote absolute frequency measurement using the intercity optical carrier fiber link between the University of Tokyo and the National Metrology Institute of Japan (NMIJ) at Tsukuba. In outline, the Sr lattice clock located in Tokyo is measured based on an H-maser (NMIJ-HM2) in Tsukuba using a fiber link, while the H-maser is calibrated using a Cs fountain clock (NMIJ-F1) and a frequency link to the TAI. The 120-km fiber network consists of a 100-km fiber network provided by the Japan Gigabit Network 2, and commercial fibers in both Tokyo (15 km) and Tsukuba (5 km). The fiber losses at 1.5 µm are 15 dB in Tokyo, 30 dB between Tokyo and Tsukuba, and 7 dB in Tsukuba.

A self-referenced [2] fiber comb is phase-locked to the H-maser. A narrow linewidth fiber DFB laser operating at 1542 nm is phase-locked to the fiber comb to convert the microwave signal of the H-maser into an optical signal for fiber transfer. The input optical power at Tsukuba is about 10 mW, which is limited by stimulated Brillouin scattering, and consequently the output power at Tokyo is about 50 nW. An external cavity diode laser located in Tokyo is used as a repeater. The laser is phase-locked to the transferred light with an offset frequency of 40 MHz to make it possible to distinguish the round-trip light from the stray reflections in the fiber. The light is then sent back to Tsukuba so that its phase can be compared with that of the original light and thus allows the detection of the fiber length fluctuation [8] caused by both acoustic noise and temperature variations. A fiber stretcher and an acousto-optic modulator are used to servo control the fiber length and thus suppress the phase noise [15].

The frequency of the repeater laser in Tokyo is doubled by using a periodically poled $LiNbO_3$ crystal, and is measured by using a self-referenced Ti:sapphire (Ti:S) comb at 771 nm. The Ti:S comb phase-locked to the Sr lattice clock converts the Sr frequency to the entire Ti:S comb measurement range covering 500 to 1100 nm. The Sr lattice clock is a one-dimensional optical lattice clock with ultracold spin-polarized fermionic $^{87}$Sr atoms [5].

Figure 2(a) shows the time variation of the measured Sr frequency based on a fiber-linked H-maser with a counter gate time of 1 s. The Allan standard deviation (fractional frequency instability) of the measured frequency is calculated and shown as a solid curve with filled circles in Fig. 2(b). The Allan standard deviation is given by $4.7 \times 10^{-13}/\tau^{1/2}$ (blue solid line), where the $1/\tau^{1/2}$ slope indicates white frequency noise. The Allan standard deviation at 1 s is mainly limited by the instability of NMIJ-HM2. The typical Allan standard deviation of the Sr lattice clock is also shown in Fig. 2(b) as a dotted line ($1 \times 10^{-14}/\tau^{1/2}$ for an averaging time $\tau > 100$ s), and is about 30 times smaller than that of the measured frequency. The frequency instability of the coherent optical carrier transfer with active phase noise cancellation is $8 \times 10^{-16}$ at 1 s and is shown as a dashed line in Fig. 2(b). This result indicates that clock's performance is transferred without degrading its stability.

The measurement yields an average frequency of $f_{ave}$=429228004229874.1 Hz. Table 1 shows the applied systematic corrections and their uncertainties. The main correction involves the frequency of NMIJ-HM2 referred to NMIJ-F1. Since we are evaluating the uncertainty of NMIJ-F1 (now $4\times10^{-15}$) (long-term instability well below $1\times10^{-15}$), and we have been reporting NMIJ-F1 to the International Bureau of Weights and Measures (BIPM) during the measurement [16], we traced the 'second' back to TAI to improve the accuracy of the present measurement. The relatively large uncertainty comes from the measurement-time-limited stability that is calculated from the Allan standard deviation of the measured frequency $4.7\times10^{-13}/\tau^{1/2}$ at 10682 s. The systematic corrections of the Sr lattice clock itself correspond to the results of our previous investigations [5], except with respect to the gravitational shift. We found a correction of + 9 m for the previous Sr clock attitude calculation, which led to a – 0.43 Hz correction in the gravitational shift. The newly determined absolute frequency is well within the uncertainty of our previous measurement [5] (as shown in Fig. 3). The overall uncertainty of the measured frequency is 2.4 Hz.

Figure 3 compares the absolute frequencies of the Sr lattice clock measured by the three groups. Our newly determined Sr frequency obtained using the optical carrier fiber link agrees with the JILA [17] and SYRTE [18] results with a standard deviation of 0.27 Hz (corresponding to a fractional uncertainty of $6\times10^{-16}$). The absolute frequency measurements of the JILA and the SYRTE Sr lattice clocks are based on their local fountains, NIST-F1 and SYRTE-FO1&FO2, respectively, which have a frequency uncertainty of about $4\times10^{-16}$. In contrast, the absolute frequency measurement of the Tokyo Sr lattice clock is based on a precise frequency link to TAI, which now has an unprecedentedly low uncertainty near $5\times10^{-16}$ obtained by averaging an ensemble of nine fountains from seven laboratories including NMIJ-F1 over 2 years [19]. The agreement between the three groups as regards the Sr frequency measurement has reached the level of the current best evaluations of Cs primary frequency standards for the second. It is worth mentioning that we can no longer expect to achieve better agreement as long as we rely on the current TAI, which is maintained by an ensemble of Cs clocks. This fact naturally suggests the need to redefine the second in the near future and thus achieve a better international clock comparison beyond the mid $10^{-16}$ fractional level, which is the Cs limit.

We acknowledge support from K. Kobayashi, K. Nakamura, T. Higo and K. Tanaka regarding the arrangement of the optical fiber link. We are grateful to Y. Nakajima, A. Onae and S. Ohshima for technical support and useful discussions. This work was supported by Japan Gigabit Network 2 (JGN2).

**References**

1. T. Udem, R. Holzwarth, and T. W. Hansch, "Optical frequency metrology," Nature **416**, 233-237 (2002).

2. D. J. Jones, S. A. Diddams, J. K. Ranka, A. Stentz, R. S. Windeler, J. L. Hall, and S. T. Cundiff, "Carrier-envelope phase control of femtosecond mode-locked lasers and direct optical frequency synthesis," Science **288**, 635-639 (2000).

3. H. Katori, "Spectroscopy of strontium atoms in the Lamb-Dicke confinement," in *Proceedings of the 6th Symposium on Frequency Standards and Metrology*, P. Gill, ed. (World Scientific, Singapore, 2002), pp. 323-330.

4. M. Takamoto, F. L. Hong, R. Higashi, and H. Katori, "An optical lattice clock," Nature **435**, 321-324 (2005).

5. M. Takamoto, F. L. Hong, R. Higashi, Y. Fujii, M. Imae, and H. Katori, "Improved frequency measurement of a one-dimensional optical lattice clock with a spin-polarized fermionic Sr-87 isotope," Journal of the Physical Society of Japan **75** (2006).

6. A. D. Ludlow, M. M. Boyd, T. Zelevinsky, S. M. Foreman, S. Blatt, M. Notcutt, T. Ido, and J. Ye, "Systematic study of the Sr-87 clock transition in an optical lattice," Physical Review Letters **96** (2006).

7. R. Le Targat, X. Baillard, M. Fouche, A. Brusch, O. Tcherbakoff, G. D. Rovera, and P. Lemonde, "Accurate optical lattice clock with Sr-87 atoms," Physical Review Letters **97** (2006).

8. L. S. Ma, P. Jungner, J. Ye, and J. L. Hall, "DELIVERING THE SAME OPTICAL FREQUENCY AT 2 PLACES - ACCURATE CANCELLATION OF PHASE NOISE INTRODUCED BY AN OPTICAL-FIBER OR OTHER TIME-VARYING PATH," Optics Letters **19**, 1777-1779 (1994).

9.      J. Ye, J. L. Peng, R. J. Jones, K. W. Holman, J. L. Hall, D. J. Jones, S. A. Diddams, J. Kitching, S. Bize, J. C. Bergquist, L. W. Hollberg, L. Robertsson, and L. S. Ma, "Delivery of high-stability optical and microwave frequency standards over an optical fiber network," Journal of the Optical Society of America B-Optical Physics **20**, 1459-1467 (2003).

10.     C. Daussy, O. Lopez, A. Amy-Klein, A. Goncharov, M. Guinet, C. Chardonnet, F. Narbonneau, M. Lours, D. Chambon, S. Bize, A. Clairon, G. Santarelli, M. E. Tobar, and A. N. Luiten, "Long-distance frequency dissemination with a resolution of 10(-17)," Physical Review Letters **94** (2005).

11.     S. M. Foreman, A. D. Ludlow, M. H. G. de Miranda, J. E. Stalnaker, S. A. Diddams, and J. Ye, "Coherent optical phase transfer over a 32-km fiber with 1 s instability at 10(-17)," Physical Review Letters **99** (2007).

12.     N. R. Newbury, P. A. Williams, and W. C. Swann, "Coherent transfer of an optical carrier over 251 km," Optics Letters **32**, 3056-3058 (2007).

13.     I. Coddington, W. C. Swann, L. Lorini, J. C. Bergquist, Y. Le Coq, C. W. Oates, Q. Quraishi, K. S. Feder, J. W. Nicholson, P. S. Westbrook, S. A. Diddams, and N. R. Newbury, "Coherent optical link over hundreds of metres and hundreds of terahertz with subfemtosecond timing jitter," Nature Photonics **1**, 283-287 (2007).

14.     A. D. Ludlow, T. Zelevinsky, G. K. Campbell, S. Blatt, M. M. Boyd, M. H. G. de Miranda, M. J. Martin, J. W. Thomsen, S. M. Foreman, J. Ye, T. M. Fortier, J. E. Stalnaker, S. A. Diddams, Y. Le Coq, Z. W. Barber, N. Poli, N. D. Lemke, K. M. Beck, and C. W. Oates, "Sr lattice clock at 1 x 10(-16) fractional uncertainty by remote optical evaluation with a Ca clock," Science **319**, 1805-1808 (2008).

15. M. Musha, K. Nakagawa, K.-i. Ueda, and F.-L. Hong, "Precision optical carrier transmission over 110km through urban fiber network," in *Proceedings of URSI General Assembly 2008*(Chicago, Illinois, 2008), p. A02.03.

16. International Bureau of Weights and Measures (BIPM), Circular T, No. 243, (March 2008), http://www.bipm.org/jsp/en/TimeFtp.jsp?TypePub=publication.

17. G. K. Campbell, A. D. Ludlow, S. Blatt, J. W. Thomsen, M. J. Martin, M. H. G. d. Miranda, T. Zelevinsky, M. M. Boyd, J. Ye, S. A. Diddams, T. P. Heavner, T. E. Parker, and S. R. Jefferts, " The absolute frequency of the $^{87}$Sr optical clock transition," arxiv.org/abs/0804.4509v1 (2008).

18. X. Baillard, M. Fouche, R. Le Targat, P. G. Westergaard, A. Lecallier, F. Chapelet, M. Abgrall, G. D. Rovera, P. Laurent, P. Rosenbusch, S. Bize, G. Santarelli, A. Clairon, P. Lemonde, G. Grosche, B. Lipphardt, and H. Schnatz, "An optical lattice clock with spin-polarized Sr-87 atoms," European Physical Journal D **48**, 11-17 (2008).

19. T. E. Parker, "A Comparison of Cesium Fountain Primary Frequency Standards," in *22nd European Frequency and Time Forum*(Toulouse, France, 2008).

**Table 1 Systematic frequency corrections and uncertainties for the Sr optical lattice clock.**

| Effect | Correction (Hz) | Uncertainty (Hz) |
|---|---|---|
| Frequency link [(NMIJ-HM2) – (NMIJ-F1)] | -54.90 | 1.15 |
| Frequency link [(NMIJ-F1) – TAI] | -1.50 | 0.69 |
| Frequency measurement | 0 | 1.9 |
| Blackbody shift | 2.4 | 0.2 |
| Second order Zeeman shift | 0.77 | 0.01 |
| Gravitational shift | -0.93 | 0.09 |
| Collision shift | 0.4 | 0.3 |
| Lattice scalar light shift | -0.22 | 0.33 |
| Lattice fourth order light shift | -0.017 | 0.015 |
| Probe laser light shift | 0.030 | 0.001 |
| Systematic total | -53.97 | 2.4 |

Figure captions

Fig. 1. (Color online) Diagram of remote absolute frequency measurement using a 120-km coherent optical transfer between two cities. Ti:s, Ti:sapphire; $f_{rep}$, repetition rate of the comb.

Fig. 2. (Color online) Measurement results for a Sr lattice clock based on a fiber-linked H-maser. (a) Time variation of the measured frequency. (b) Allan standard deviation of the measured frequency (solid curve with filled circles). Also shown are the Allan standard deviations of the Sr optical lattice clock (dotted line) and the fiber link (dashed line).

Fig. 3. (Color online) Comparison of the absolute frequencies of Sr lattice clocks measured by three groups. The absolute frequency of the Sr lattice clock in Tokyo is 429228004229874.1(2.4) Hz, which agrees with the results from JILA [17] and SYRTE [18] with a fractional uncertainty of $6 \times 10^{-16}$. Also shown is the result of our last measurement [5].

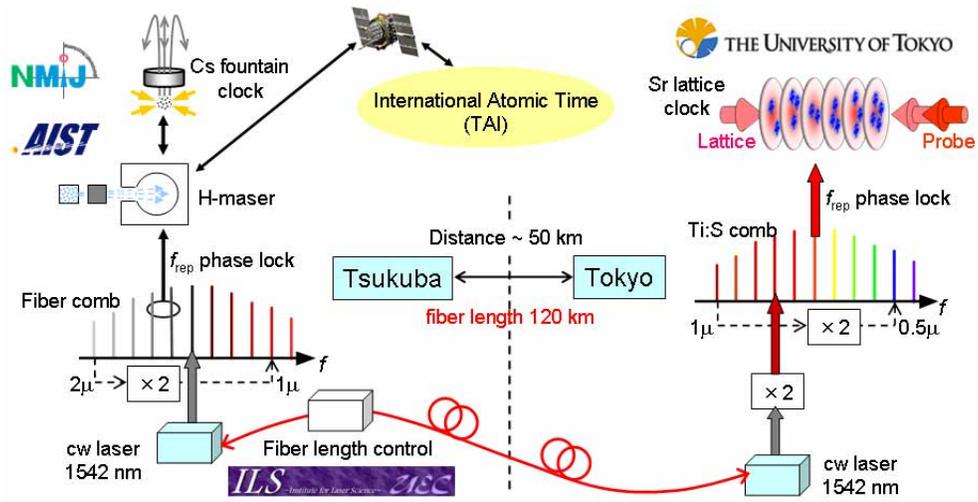

Fig. 1

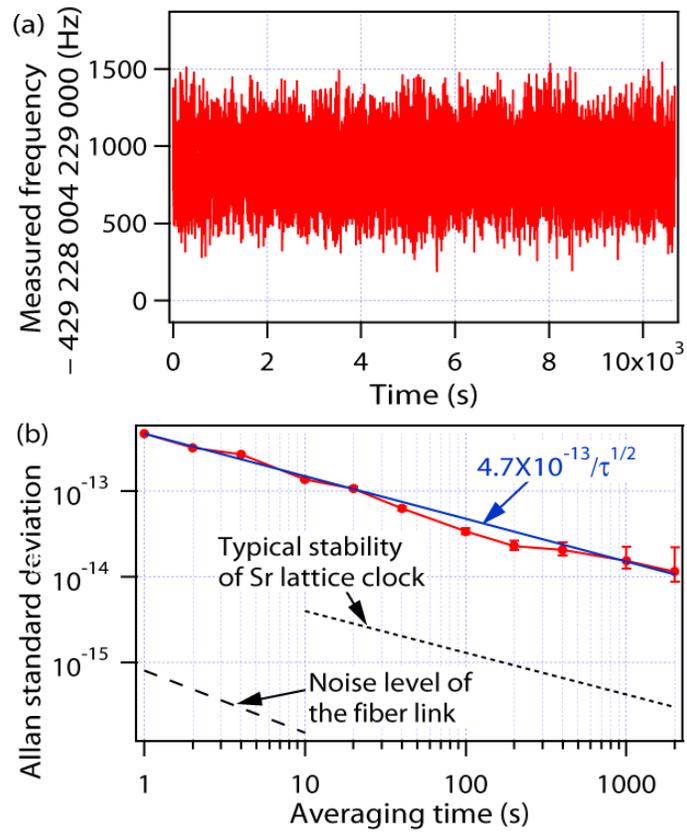

Fig. 2

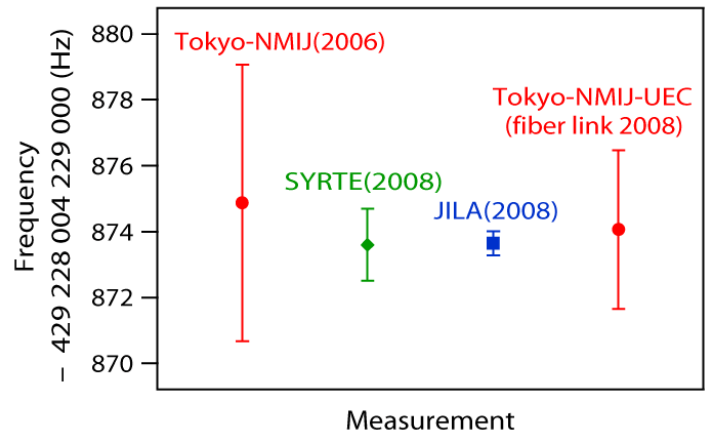

Fig. 3